\documentclass[american,aip,jap,reprint]{revtex4-1}
\usepackage[T1]{fontenc}
\usepackage[utf8]{inputenc}
\usepackage{color}
\usepackage{babel}
\usepackage{amsmath}
\usepackage{graphicx}
\usepackage[unicode=true,
 bookmarks=true,bookmarksnumbered=false,bookmarksopen=false,
 breaklinks=true,pdfborder={0 0 0},pdfborderstyle={},backref=false,colorlinks=true]
 {hyperref}
\hypersetup{pdftitle={First-principles calculations of iron-hydrogen reactions in silicon },
 pdfauthor={Paulo Santos, José Coutinho, Sven Öberg},
 allcolors=blue}
\usepackage{breakurl}

\makeatletter

\newcommand{\lyxmathsym}[1]{\ifmmode\begingroup\def\b@ld{bold}
  \text{\ifx\math@version\b@ld\bfseries\fi#1}\endgroup\else#1\fi}

\providecommand{\tabularnewline}{\\}

\usepackage{booktabs}
\usepackage{amsmath}

\makeatother

\begin{document}

\title{First-principles calculations of iron-hydrogen reactions in silicon }

\author{Paulo Santos}
\email{paulodsantos@ua.pt}

\affiliation{Department of Physics and I3N, University of Aveiro, Campus Santiago,
3810-193 Aveiro, Portugal}

\author{José Coutinho}

\affiliation{Department of Physics and I3N, University of Aveiro, Campus Santiago,
3810-193 Aveiro, Portugal}

\author{Sven Öberg}

\affiliation{Department of Engineering Sciences and Mathematics, Luleå University
of Technology, Luleå S-97187, Sweden}
\begin{abstract}
Controlling the contamination of silicon materials by iron, especially
dissolved interstitial iron (Fe$_{\mathrm{i}}$), is a longstanding
problem with recent developments and several open issues. Among these
we have the question whether hydrogen can assist iron diffusion, or
if significant amounts of substitutional iron (Fe$_{\mathrm{s}}$)
can be created. Using density functional calculations we explore the
structure, formation energies, binding energies, migration, and electronic
levels of several FeH complexes in Si. We find that a weakly bound
Fe$_{\mathrm{i}}$H pair has a migration barrier close to that of
isolated Fe$_{\mathrm{i}}$ and a donor level at $E_{\mathrm{v}}+0.5$~eV.
Conversely, Fe$_{\mathrm{i}}$H$_{2}(0/+)$ is estimated at $E_{\mathrm{v}}+0.33$~eV.
These findings suggest that the hole trap at $E_{\mathrm{v}}+0.32$~eV
obtained by capacitance measurements should be assigned to Fe$_{\mathrm{i}}$H$_{2}$. Fe$_{\mathrm{s}}$H-related complexes show only deep acceptor activity
and are expected to have little effect on minority carrier life-time
in $p$-type Si. The opposite conclusion can be drawn for $n$-type
Si. We find that while in H-free material Fe$_{\mathrm{i}}$ defects
have lower formation energy than Fe$_{\mathrm{s}}$, in hydrogenated
samples Fe$_{\mathrm{s}}$-related defects become considerably more
stable. This would explain the observation of an EPR signal attributed
to a Fe$_{\mathrm{s}}$H-related complex in hydrogenated Si, which
was quenched from above 1000$^{\circ}$C to iced-water temperature.
\emph{Published in Journal of Applied Physics}\textbf{\emph{123}}\emph{,
245703 (2018).} \texttt{\textbf{\textcolor{blue}{https://doi.org/10.1063/1.5039647}}}
\end{abstract}
\maketitle

\section{Introduction}

Iron is a fundamental constituent of many tools and industrial equipments,
it is present in silicon raw materials, and that makes Fe contamination
of Si ingots virtually unavoidable.\citep{Jastrzebski1994,Istratov2000}
Stringent control of Fe impurities in Si is particularly critical
in electronic- and solar-grade materials, as a donor level at $E_{\mathrm{v}}+0.38$~eV
from interstitial iron (Fe$\mathrm{_{i}}$),\citep{Feichtinger1978}
leads to powerful minority carrier recombination activity in $p$-type
Si,\citep{Istratov1999} most often the material of choice for the
fabrication of Si solar cells.

Hydrogenation of Silicon wafers has been recurrently applied in order
to passivate, or at least reduce, the recombination activity from
several defects and contaminants, including iron and other transition
metals.\citep{Pearton1982,Singh1986,Istratov2000,Peaker2012,Liu2014,Scheffler2015,Mullins2017}
In Si photovoltaics this hydrogenation process is usually accomplished
by means of depositing and firing a hydrogen-soaked SiN$_{x}$ layer
on top of Si, which besides the surface-passivation effect, it also
works as an anti-reflection coating for the front surface of the cell.\citep{Jiang2003,Holt2003}
Other types of hydrogen introduction for passivation treatments have
also been considered, including proton implantation\citep{Kouketsu1996}
or H-plasma exposure,\citep{Pearton1984} but none is as convenient
as the nitridation process.

The interaction between Fe and H in silicon has been addressed in
the past. Early studies by Pearton and Tavendale\citep{Pearton1984}
reported the passivation of iron- and silver-related centers in $p$-type
samples exposed to Hydrogen plasma. Here, the introduction of the
metallic impurities was accomplished by high-temperature evaporation,
and from deep-level transient spectroscopy (DLTS), the suppression
of an electrical level at $E_{\mathrm{v}}+0.32$~eV, by the time
connected to a Fe-O complex,\citep{Wunstel1981} after the H-plasma
treatment was announced as an interaction between H and Fe. No direct
interaction between H and interstitial Fe (Fe$_{\mathrm{i}}$) was
reported on these studies. More than a decade later, thermally stimulated
capacitance (TSCAP) measurements performed by Sadoh \emph{et~al.}\citep{Sadoh1997}
using iron doped floating-zone $n$\nobreakdash-type Si that was
subject to wet-etching, displayed the same level, which was then reassigned
to a FeH complex. Annealing studies showed that after 30-minute treatments
at $175\lyxmathsym{º}\mathrm{C}$, the level disappeared, suggesting
a low binding energy between Fe and H species. Despite these conclusions,
there was no direct evidence for the presence of either Fe or H in
the `$E_{\mathrm{v}}+0.32$~eV' center.

More recently, further evidence for a Fe-H complex in $p$-type silicon
was reported by Leonard and co-workers.\citep{Leonard2015} The hydrogen
was introduced into the samples from a silicon nitride layer grown
by plasma enhanced chemical vapor deposition. After a reverse-bias
annealing at about 100$^{\circ}$C, a hole trap was observed in the
DLTS spectrum from samples with high concentrations of Fe and H. This
trap was only stable up to 125$^{\circ}$C and it was related to a
donor level at $E_{\mathrm{v}}+0.31$~eV. The electronic signature
of this trap coincided with that reported in Ref.~\onlinecite{Sadoh1997}.

The interaction of hydrogen with iron-related defects such as the
iron-boron (FeB) pair and Fe$_{\mathrm{i}}$ itself was also addressed
by Kouketsu and Isomae.\citep{Kouketsu1996} In this case, hydrogen
was introduced by proton implantation, leading different observations
when compared to those where hydrogen plasma was used. Accordingly,
along with the disappearance of the DLTS signals related to Fe$_{\mathrm{i}}$
and FeB, the emergence of two hole traps at $E_{\mathrm{v}}+0.23$~eV
and $E_{\mathrm{v}}+0.38$~eV was observed. It was not clear though,
whether these new traps arose from the reaction between H and Fe$_{\mathrm{i}}$-related
complexes, or on the other hand, from complexes involving H and intrinsic
defects (resulting from the implantation damage), eventually combined
with Fe.

The interaction of H with FeB pairs was also studied by Yakimov and
Parakhonsky using wet-etching,\citep{Yakimov1997} therefore avoiding
implantation damage effects. Their results suggested that at room-temperature,
the introduction of hydrogen actually leads to dissociation of the
FeB pairs in the near surface layer, thus increasing the concentration
of interstitial iron. The process leading to such dissociation is
not well understood, and therefore it is worth of further investigation.

Electron paramagnetic resonance (EPR) was also employed in the study
of H-passivation of Fe in Si. Accordingly, the existence of a FeH
complex with a binding energy of $\sim1.3$~eV and stable up to<
$T=220^{\circ}$C was inferred after comparing EPR spectra from Fe-doped
and (Fe,H)-co-doped $n$-type floating zone Si.\citep{Takakashi1999}
Interestingly, the EPR signal which was assigned to the FeH complex
was isotropic ($T_{d}$ symmetry). Unfortunately, the authors could
not unambiguously demonstrate the presence of H atoms in the center,
either from the EPR data or from local vibrational mode absorption
using deuterated samples.

Theoretical work by the Estreicher group proposed\emph{ }two stable
structures for the FeH complex, namely Fe$_{\mathrm{i}}$H and Fe$_{\mathrm{s}}$H,
involving interstitial and substitutional Fe, respectively.\citep{Szwacki2008}
Although under equilibrium conditions iron impurities are located
at interstitial sites, there is evidence for the existence of substantial
concentrations of iron substitutional (Fe$_{\mathrm{s}}$) provided
by emission channeling\citep{Wahl2005,Silva2013} and Mössbauer spectroscopy.\citep{Gilles1990,Langouche1992,Weyer1997,Weyer1999,Yoshida2002,Yoshida2003}
The Fe$_{\mathrm{i}}$H model consists on a Fe-H dimer with trigonal
symmetry, with the Fe atom placed at the hexagonal interstitial site,
while H is located close to a neighboring tetrahedral interstitial
site. This structure was anticipated to produce two electrical levels,
a donor at $E\mathrm{_{v}+}0.36\,\mathrm{eV}$ and an acceptor at
$E\mathrm{_{c}-}0.26\,\mathrm{eV}$.\citep{Szwacki2008} The authors
also estimated an energy gain of $0.82$~eV for the reaction $\mathrm{Fe_{i}}+\mathrm{H_{BC}}\rightarrow\mathrm{Fe_{i}H}$
(assuming a bond-centered configuration for hydrogen and neutral defects
only), consistent with the thermal stability of the `$E_{\mathrm{v}}+0.32$~eV'
trap (annealing temperature of about 175$^{\circ}$C) as reported
by Sadoh \emph{et~al.}\citep{Sadoh1997} This reaction is expected
to be hindered in \emph{p}-type Si due to electrostatic repulsion
between $\mathrm{Fe_{i}^{+}}$ and $\mathrm{H}^{+}$ ions.

The second structure, $\mathrm{Fe_{s}H}$, was found to comprise an
iron atom locked at a substitutional site, with the hydrogen atom
roaming almost freely around it. This model was assigned to the EPR
spectra reported by Takakashi and co-workers,\citep{Takakashi1999}
conforming to the observed isotropic symmetry. $\mathrm{Fe_{s}H}$
was predicted to produce an acceptor level at $E\mathrm{_{c}-}0.62\,\mathrm{eV}$.
The estimated binding energy for this structure was approximately
$1.4\,\mathrm{eV}$, also in good agreement with the 1.3~eV binding
energy estimated from the ESR measurements\citep{Takakashi1999}.
Despite the agreement, the EPR data was acquired in $n$-type material
and for these conditions the proposed Fe$_{\mathrm{s}}$H complex
would be in a diamagnetic negative charge state, raising doubts to
the correctness of this assignment.

While the early literature indicates a relatively low thermal stability
for Fe-H complexes (the measured binding energies and annealing
experiments on one hand,\citep{Sadoh1997,Takakashi1999,Leonard2015}
and the first-principles calculations by Szwacki \emph{et~al.}\citep{Szwacki2008}
on the other hand, suggest that these complexes can only survive to
temperatures of at most $125\textrm{-}200\lyxmathsym{º}\mathrm{C}$), 
more recently the solar-Si community has turned the attention
to the effect of hydrogenation on Fe diffusivity and gettering at
higher temperatures. For instance, Ref.~\onlinecite{Karzel2013}
reports a prominent decrease in the concentration of Fe$\mathrm{_{i}}$
after exposing multicrystalline Si wafers to a microwave-induced remote
hydrogen plasma, followed by H-effusion during 300-500$^{\circ}$C-anneals.
This is a surprisingly stable process (when compared to H-passivation
using etched samples), and it was tentatively explained as a consequence
of an enhanced diffusion of Fe$\mathrm{_{i}}$ caused by the introduction
of H, thus leading to a faster gettering kinetics. 

On the other hand, Liu \emph{et~al.}\citep{Liu2014} questioned the
picture of a H-enhanced diffusivity of Fe$_{\mathrm{i}}$, arguing
that if that was the case, any increase of the annealing temperatures
should lead to the observation of further gettering of Fe. However,
they report that at $700\,\lyxmathsym{º}\mathrm{C}$, the amount
of dissolved iron reaches a minimum of about 1\% of the original content,
and recovers at higher temperatures. It was then suggested that up
to $700\,\lyxmathsym{º}\mathrm{C}$, the formation of a Fe-H complex
should account for the decrease of Fe$_{\mathrm{i}}$, and above that
temperature the dissociation of the complex becomes dominant. This
view was latter revised after secondary-ion mass spectroscopy (SIMS)
measurements combined with DLTS, annealing and analysis of the iron-decay
kinetics.\citep{Liu2016} From the observed accumulation of Fe at
the SiN$_{x}$ capping layer, it was concluded that the iron reduction
in the Si bulk takes place via gettering at the silicon nitride films.
Hydrogenation of Fe$_{\mathrm{i}}$ at high-temperatures was ruled
out based on the lack of electrical activity in the Si as monitored
by DLTS. Further, upon removal of the SiN$_{x}$, subsequent high-temperature
anneals did not reveal any electrical activity either.

With these observations in mind, we endeavored to calculate the stability,
electrical activity and migration ability of FeH-related complexes.
After describing the methodology, we report on the atomistic structure
and energetics of FeH defects, their electronic activity, thermal
stability and hydrogen-assisted migration of interstitial iron. We
end up with a discussion of the results and conclusions.

\section{Method}

We employed the \emph{Vienna Ab-initio Simulation Package} (VASP)
code\citep{Kresse1993,Kresse1994,Kresse1996} to perform density functional
calculations concerning the relative stability, formation energies,
electrical levels and migration barriers of FeH complexes in Si. These
calculations were based on the projector-augmented wave (PAW) \citep{bloch1994}
method using the generalized gradient approximated exchange-correlation
functional of Perdew, Burke and Ernzerhof.\citep{Perdew1996} The
PAW potentials for Fe, Si and H species were generated in the $3\mathrm{s}^{2}3\mathrm{p}^{6}3\mathrm{d}^{7}4\mathrm{s}^{1}$,
$3\mathrm{s}^{2}3\mathrm{p}^{2}$and $1\mathrm{s}^{1}$ valence configurations,
respectively. The Kohn-Sham states were expanded in plane-waves with
a cut off energy of $450\,\mathrm{eV}$. 

Our atomistic models for the defects under scrutiny were inserted
on 216 Si atom supercells with a theoretical lattice parameter of
$a_{0}=5.456$~Å. We employed a $2\times2\times2$ Monkhorst and
Pack $\mathbf{k}$-point grid to sample the Brillouin zone.\citep{Monkhorst1976}
The structural optimization of our defect models was done through
a conjugate gradient method, with a convergence threshold of $2.5\times10^{-3}$~eV/Å
for the maximum force acting on the nuclei. The self-consistent electronic
relaxation cycles were computed with an accuracy of $10^{-7}\,\mathrm{eV}$.

We employed the marker method \citep{Resende1999,Coutinho2003} to
assess the electrical activity of the FeH complexes. The markers for
the double acceptor, acceptor and donor levels are, respectively:
$\mathrm{Ni_{s}(=/-)=E_{c}-0.08\,\mathrm{eV}}$, $\mathrm{Ni_{s}(-/0)=E_{c}-0.31\,\mathrm{eV}}$
and $\mathrm{Fe_{i}(0/+)=E_{v}+0.38\,\mathrm{eV}}$.\citep{Shiraishi1999,Feichtinger1978}
The respective electron affinities ($A$) and ionization potentials
($I$) calculated using the same 216-Si supercells are: $A\mathrm{\left\{ Ni_{s}(=/-)\right\} =6.06\,\mathrm{eV}}$,
$A\mathrm{\left\{ Ni_{s}(-/0)\right\} =5.99\,\mathrm{eV}}$ and $I\mathrm{\left\{ Fe_{i}(0/+)\right\} =5.73}\,\mathrm{eV}$.
The image-charge corrections for all the markers and defects under
scrutiny were accounted for using the algorithm proposed by Freysoldt,
Neugebauer and Van de Walle.\citep{Freysoldt2009} The marker method
consists on a direct comparison between ionization potentials (or
electron affinities) of the marker and that of the defect under scrutiny.
While charge-corrections to the energies are of the order of hundreds
of meV, these essentially cancel in the calculated electronic levels,
becoming a few meV. The error bar of the calculated levels was estimated
at about 0.1~eV. This figure was estimated by calculating the Fe$_{\mathrm{i}}$
and Fe$_{\mathrm{s}}$ levels, but instead of a defect marker, ionization
energies and electron affinities of a bulk supercell were assumed
as reference energies for the valence band top and conduction band
bottom edges, respectively.

Formation energies of neutral defects, $E_{\mathrm{f}}^{0}$ , where
determined using the following expression:

\begin{equation}
E_{\mathrm{f}}^{0}=E_{\mathrm{def}}^{0}-\sum_{i}n_{i}\mu_{i},\label{eq:Eform}
\end{equation}
where $E_{\mathrm{def}}^{0}$ stands for the total energy of a neutral
defective supercell made of $n_{i}$ atoms of species $i$ with chemical
potential $\mu_{i}$. Chemical potentials $\mu_{\mathrm{Si}}=-5.42$~eV,
$\mu_{\mathrm{Fe}}=-9.68$~eV and $\mu_{\mathrm{H}}=-3.39$~eV were
obtained from bulk Si, iron disilicide ($\mathrm{\beta\textrm{-}FeSi_{2}}$)
and a H$_{2}$ molecule in a box, respectively. The formation energy
of a defect in charge state $q$ has a $E_{\mathrm{f}}^{q}\sim qE_{\mathrm{F}}$
dependence, where $E_{\mathrm{F}}$ is the Fermi energy with respect
to the valence band top. The calculation of $E_{\mathrm{f}}^{q}$
was carried out combining Eq.~\ref{eq:Eform} and the results from
the marker method described above. See Ref.~\onlinecite{Coutinho2012}
for further details.

Hydrogen-assisted migration of interstitial iron was also investigated.
We employed a 7-image nudged elastic band (NEB) method\citep{Henkelman2000}
in order to estimate the migration/transformation barriers of FeH-related
defects in the neutral and positive charge states. These are the relevant
states to be considered in $p$-type material.

\section{Results }

\subsection{Defect Structures}

In order to determine the ground state structures of Fe$_{\mathrm{i}}$H
and Fe$_{\mathrm{s}}$H defects, we started from the relaxed structures
of Fe$_{\mathrm{i}}$ and Fe$_{\mathrm{s}}$, respectively, and introduced
one or two hydrogen atoms at several possible sites, either bonding
directly to the Fe atom along different directions, next to their
silicon first neighbors, at second-neighboring Si-Si bond-center sites,
or near the Fe\nobreakdash-Si bond-center site.

\begin{figure}
\includegraphics[width=7cm]{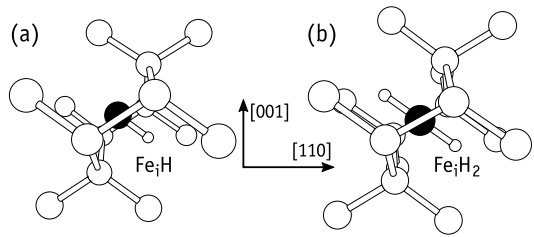}\caption{\label{fig1}Ground state structure of (a) $\mathrm{\mathrm{Fe_{i}H}}$
and (b) $\mathrm{\mathrm{Fe_{i}H_{2}}}$ defects in silicon. The large
white, black and small white spheres represent Si, Fe and H atoms,
respectively. Fe-H bonds are $1.47$~Å long on both defects.}
\end{figure}

In line with the calculations reported by Szwaki \emph{et~al.},\citep{Szwacki2008}
we predict that after trapping one hydrogen atom, the Fe$_{\mathrm{i}}$
atom becomes more stable near the hexagonal site while connecting
to H along the trigonal axis. The resulting Fe$_{\mathrm{i}}$H structure
is depicted in Fig.~\ref{fig1}(a). The defect is stable in the negative,
neutral and positive charge states with spin 0, 1/2 and 1, respectively.
The structure of the Fe$_{\mathrm{i}}$H$_{2}$ complex is analogous,
with both H atoms bonded to Fe$_{\mathrm{i}}$ pointing towards opposite
directions along a common $\langle111\rangle$ axis. This is shown
in Fig.~\ref{fig1}(b), and the defect is also stable in the $-$,
$0$, and $+$ charge states with spin 1/2, 1 and 1/2, respectively. 

\begin{figure}
\includegraphics[width=6cm]{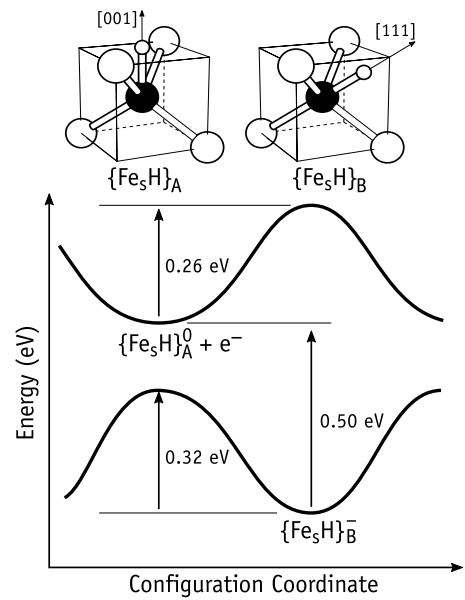}\caption{\label{fig2}Configuration coordinate diagram of the $\mathrm{Fe_{s}H}$
defect. The upper part of the figure includes the two ground-state
configurations $\mathrm{\{Fe_{s}H\}_{A}}$ and $\mathrm{\{Fe_{s}H\}_{B}}$
for neutral and negative charge states, respectively. The black sphere
represents Fe while the smaller and larger white spheres represent
H and Si, respectively.}
\end{figure}

In the case of $\mathrm{Fe_{s}H}$, we found it to be bistable, with
H connecting to Fe$_{\mathrm{s}}$ along the $\langle100\rangle$
direction in the neutral charge state, or along $\langle111\rangle$
towards the tetrahedral interstitial site in the negative charge state.
These structures are represented in the configuration coordinate diagram
of Fig.~\ref{fig2}, and are labeled as $\mathrm{\{Fe_{s}H\}_{A}^{0}}$and
$\mathrm{\{Fe_{s}H\}_{B}^{-}}$ respectively. They show spin 1/2 and
0, respectively, and each of them has a unique minimum in the potential
energy surface: for the negatively charged defect, structural optimization
initiated in the $\mathrm{\{Fe_{s}H\}_{A}}$ configuration relaxed
into $\mathrm{\{Fe_{s}H\}_{B}}$. Conversely, neutral $\mathrm{\{Fe_{s}H\}_{B}}$
is unstable and spontaneously relaxes to $\mathrm{\{Fe_{s}H\}_{A}^{0}}$.
The energy barriers for conversion between $\mathrm{\{Fe_{s}H\}_{A}}\leftrightarrow\mathrm{\{Fe_{s}H\}_{B}}$
were calculated as 0.26~eV and 0.32~eV for neutral and negative
charged defects. These figures are at variance with those obtained
in Ref.~\onlinecite{Szwacki2008} (0.08~eV) and we can only suggest
that in that work the Brillouin-zone sampling that was employed ($\Gamma$-only)
was not sufficiently dense considering the size of the supercells
(64 atoms).

The capture of a second hydrogen atom by Fe$_{\mathrm{s}}$ also leads
to two stable $\mathrm{Fe_{s}H_{2}}$ configurations in different
charge states. For the neutral charge state we obtained a linear H-Fe-H
configuration with both Fe-H bonds along the $\langle100\rangle$
crystallographic axis, pointing towards opposite directions. This
structure has high symmetry ($D_{2d}$ point group) and from inspection
of the Kohn-Sham band structure with identified an empty double degenerate
state in the upper half of the gap. We label this structure $\mathrm{\{Fe_{s}H_{2}\}_{A}}$.

In the negative charge state the doublet state becomes partially populated
and the structure undergoes a Jahn-Teller distortion. This translates
into a $E_{\mathrm{JT}}\sim0.3$~eV relaxation energy and to the
formation of a slanted Fe-H bond about 10$^{\circ}$ away from the
$\langle100\rangle$ axis. Several other low-energy distortions were
found within 30~meV from the ground state. These consisted of pairs
of Fe-H bonds oriented close to $\langle100\rangle$ and $\langle111\rangle$,
like in $\mathrm{\{Fe_{s}H\}_{A}}$ and $\mathrm{\{Fe_{s}H\}_{B}}$,
respectively. All non-linear H-Fe-H defects (including the ground
state) were found with spin 1/2, and because they are all nearly degenerate,
we refer to them as $\mathrm{\{Fe_{s}H_{2}\}_{B}}$. Interestingly,
the transformation barrier between these low-symmetry structures was
found to be about 50~meV, indicating that the H atoms can roam almost
freely around the Fe$_{\mathrm{s}}$ impurity, even at cryogenic temperatures.

In the double negative charge state, the Jahn-Teller distortion becomes
stronger and both Fe$_{\mathrm{s}}$\nobreakdash-H bonds pointed
approximately along orthogonal $\langle100\rangle$ directions. The
resulting diamagnetic $\mathrm{\{Fe_{s}H_{2}\}_{B}^{=}}$ state was
found more stable than the paramagnetic $\mathrm{\{Fe_{s}H_{2}\}_{A}^{=}}$
linear structure with spin-1 by 0.29~eV. The reorientation of a Fe$_{\mathrm{s}}$\nobreakdash-H
bond now involves surmounting a barrier of 0.12~eV. All Fe-H bond
lengths in Fe$_{\mathrm{s}}$H$_{n}$ defects were in the range of
1.52-1.55~Å.

Besides the electronic energy (enthalpy), at high temperatures the
entropy may impact on the relative stability of defects, and here
the most important contributions are from configurational and vibrational
degrees of freedom. Sanati and Estreicher found that vibrational entropy
can stabilize light impurities in silicon by $\sim0.1\textrm{-}0.2$~eV
after ramping up the temperature by up to $\Delta T=800$~K.\citep{Sanati2003}
Assuming that the magnitude of Fe-H vibrational frequencies from $\mathrm{\{Fe_{s}H\}_{A}}$
and $\mathrm{\{Fe_{s}H\}_{B}}$ are not very different, the vibrational
entropy difference is expected to be minute. Configurational entropy
is normally not considered because it is difficult to estimate analytically
except for very simple structures.\citep{Estreicher2004} Within the
Boltzmann framework, the configurational entropy difference between
structures $\mathrm{\{Fe_{s}H\}_{A}}$ and $\mathrm{\{Fe_{s}H\}_{B}}$
is about $k_{\mathrm{B}}\ln(6/4)\approx3.5\times10^{-5}$~eV/K, which
translates into a mere 27~meV for $\Delta T=800$~K. Although a
detailed account of entropy effects is outside the scope of this work,
the picture just presented should not be generalized, and we should
bear in mind that entropy could have a stronger influence, particularly
when comparing energies of defects with different stoichiometry and
lattice locations.

\subsection{Electronic levels}

The donor level of interstitial iron in silicon has been experimentally
determined at $E_{\mathrm{v}}+0.38$~eV.\citep{Feichtinger1978}
We investigated the reaction of H with Fe$_{\mathrm{i}}$ in terms
of the resulting electronic activity. Since Fe$_{\mathrm{i}}$ is
displaced from the tetrahedral site to the hexagonal site upon bonding
with one or two H atoms, the electronic activity of Fe$_{\mathrm{i}}$H
complexes is expected to differ significantly from that of isolated
Fe$_{\mathrm{i}}$. In fact, we found that besides donor activity,
both Fe$_{\mathrm{i}}$H and Fe$_{\mathrm{i}}$H$_{2}$ complexes
are acceptors. For the $\mathrm{\mathrm{Fe_{i}H}}$ pair we obtain
$(-/0)$ and $(0/+)$ levels at $E\mathrm{_{c}-}0.22$~eV and $E\mathrm{_{v}+}0.50$~eV.
On the other hand, for $\mathrm{\mathrm{Fe_{i}H}}_{\mathrm{2}}$ we
calculated $(-/0)$ and $(0/+)$ levels at $E\mathrm{_{c}-}0.29$~eV
and $E\mathrm{_{v}+}0.33$~eV, respectively. No further levels were
found for Fe$_{\mathrm{i}}$H$_{n}$ defects. Although our results
for $\mathrm{\mathrm{Fe_{i}H}}$ are not far from previous theoretical
reports,\citep{Szwacki2008} the $\mathrm{\mathrm{Fe_{i}H}}(0/+)$
level seems too deep to be connected to the `$E_{\mathrm{v}}+0.32$~eV'
trap of Ref.~\onlinecite{Sadoh1997}. Alternatively, $\mathrm{\mathrm{Fe_{i}H}}_{\mathrm{2}}$
shows a $(0/+)$ transition at $E\mathrm{_{v}+}0.33$~eV, \emph{i.e.}
about the right placement within the gap, and therefore, must be considered
as potentially accountable for the above trap as well. We will come
back to this issue in Section~\ref{subsec:diffusion}.

Now we turn to the interactions between H and substitutional iron.
For $\mathrm{\{Fe_{s}H\}_{A}}$ and $\mathrm{\{Fe_{s}H\}_{B}}$ we
obtained \emph{vertical} $(-/0)$ transitions at $\mathrm{E_{c}-}0.18\,\mathrm{eV}$
and $\mathrm{E_{c}-}0.76\,\mathrm{eV}$. However, we note that $\mathrm{\{Fe_{s}H\}_{A}^{-}}$
and $\mathrm{\{Fe_{s}H\}_{B}^{0}}$ are unstable, and the relevant
thermodynamic acceptor level of Fe$_{\mathrm{s}}$H must be calculated
from ground state energies. Hence, we obtain Fe$_{\mathrm{s}}$H$(-/0)$
at $E_{\mathrm{c}}-0.50$~eV (see Fig.~\ref{fig2}). Previous first-principles
calculation\citep{Szwacki2008} assigned Fe$_{\mathrm{s}}$H in the
neutral charge state to an isotropic spin-1/2 EPR spectrum observed
in $n$-type material at a temperature as low as 10~K.\citep{Takakashi1999}
The location of the Fe$_{\mathrm{s}}$H$(-/0)$ level implies that
under these conditions, the defect would be found in a diamagnetic
negative charge state, and therefore undetectable by EPR. We could
not find a second acceptor level for Fe$_{\mathrm{s}}$H. Hence, the
assignment of the EPR data should be revised and further work is needed
to clarify this point.

For Fe$_{\mathrm{s}}$H$_{2}$ we anticipate first and second acceptor
levels at $\mathrm{E_{c}-}0.21\,\mathrm{eV}$ and $\mathrm{E_{c}-}0.30\,\mathrm{eV}$,
respectively. It is noteworthy that the second electron trap is deeper
than the first, \emph{i.e.} Fe$_{\mathrm{s}}$H$_{2}$ shows an inverted
ordering of the acceptor levels. This is commonly referred to as negative-$U$
and arrises from a strong relaxation energy along the capture sequence,
which surmounts the Coulomb repulsion between both captured electrons.
Accordingly, in the neutral charge state Fe$_{\mathrm{s}}$H$_{2}$
adopts structure A. This structure can capture a free-electron with
a binding energy of 0.21~eV. After trapping the first electron, the
structure quickly changes to $\mathrm{\{Fe_{s}H_{2}\}_{B}}$, where
some of the relaxation energy is effectively converted to an increase
of the Coulomb attraction for the second electron, leading to a binding
energy of 0.30~eV. The consequence of the negative\nobreakdash-$U$
ordering of levels is that, under equilibrium conditions, it is energetically
favorable to form a pair of $\mathrm{\{Fe_{s}H_{2}\}^{0}}$ and $\mathrm{\{Fe_{s}H_{2}\}^{=}}$
states than two $\mathrm{\{Fe_{s}H_{2}\}^{-}}$ structures (irrespectively
of the Fermi level position). Hence, Fe$_{\mathrm{s}}$H$_{2}$ has
an $(=/0)$ occupancy level that is located half-way between the first
and second acceptor levels, \emph{i.e.} Fe$_{\mathrm{s}}$H$_{2}$$(=\!/0)=E_{\mathrm{c}}-0.26$~eV.
All calculated electrical levels are shown in Table~\ref{tab1}. 

\begin{table}
\caption{\label{tab1}Calculated electrical levels for $\mathrm{\mathrm{Fe_{i}H_{n}}}$
and $\mathrm{\mathrm{Fe_{s}H_{n}}}$ complexes. All reported values
are in eV. The inverted order of levels for Fe$_{\mathrm{s}}$H$_{2}$
leads to a $(=\!/0)$ occupancy level at $E_{\mathrm{c}}-0.26$~eV
(see text).}

\begin{ruledtabular}
\begin{tabular}{cccc}
 & $E_{\mathrm{c}}\!-\!E(=\!/-)$ & $E_{\mathrm{c}}\!-\!E(-/0)$ & $E(0/+)\!-\!E_{\mathrm{v}}$\tabularnewline
\hline 
Fe$_{\mathrm{i}}$H &  & 0.22 & 0.50\tabularnewline
Fe$_{\mathrm{i}}$H$_{2}$ &  & 0.29 & 0.33\tabularnewline
Fe$_{\mathrm{s}}$H &  & 0.50 & \tabularnewline
Fe$_{\mathrm{s}}$H$_{2}$ & 0.30 & 0.21 & \tabularnewline
\end{tabular}
\end{ruledtabular}

\end{table}

\subsection{Binding energies and doping effects\label{subsec:binding}}

\begin{figure*}
\includegraphics[width=14cm]{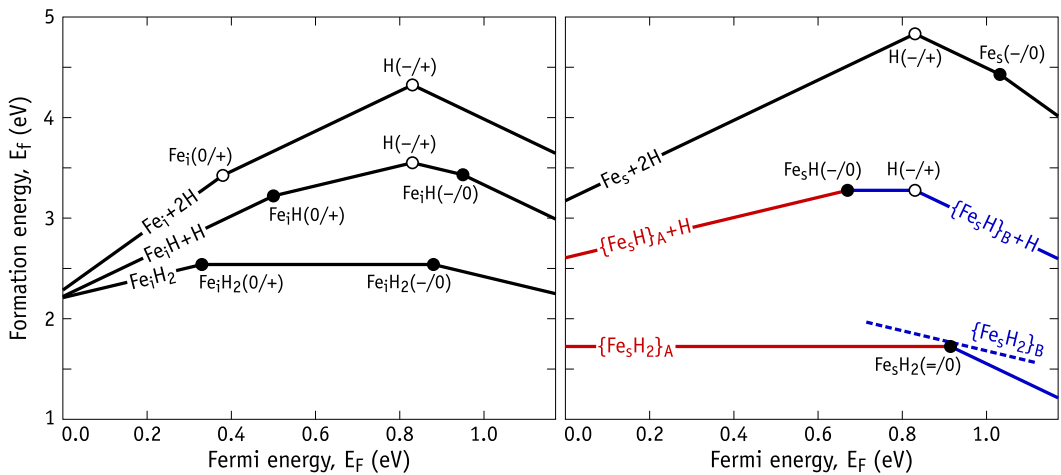}\caption{\label{fig3}Diagrams with the formation energy ($E_{\mathrm{f}}$)
as a function of the Fermi energy ($E_{\mathrm{F}}$) for different
defect arrangements involving the hydrogenation of $\mathrm{Fe_{i}}$
(left) and $\mathrm{Fe_{s}}$ (right) defects. Open and closed circles
highlight experimental and calculated electronic levels. Colors on
the right diagram are used to clarify A (red) and B (blue) structures
of Fe$_{\mathrm{s}}$H and Fe$_{\mathrm{s}}$H$_{2}$ complexes. See
text at the beginning of Section~\ref{subsec:binding} for further
details. }
\end{figure*}

The diagrams presented in Figure~\ref{fig3} show formation energies
of several defect sets involving one Fe$_{\mathrm{i}}$ impurity plus
two interstitial H atoms (left), in comparison with one Fe$_{\mathrm{s}}$
impurity plus two interstitial H atoms (right) as a function of the
Fermi energy. On each diagram the stoichiometry is conserved. The
formation energy (vertical) scales are identical for a convenient
comparison. Each sequence of connected segments relates to a particular
set involving a Fe$_{\mathrm{i}}$H$_{n}$ or Fe$_{\mathrm{s}}$H$_{n}$complex
plus $2-n$ remote interstitial H atoms. The formation energy is proportional
to $qE_{\mathrm{F}}$, with $q$ and $E_{\mathrm{F}}$ being the net
charge of the whole defect set and Fermi level, respectively. Hence,
positive-, zero- and negative-sloped segments refer to defect sets
with net positive, neutral and negative charge, respectively. For
instance, for Fe$_{\mathrm{i}}$H plus a remote H atom (on the left
diagram), as the Fermi energy goes from the valence band top ($E_{\mathrm{F}}=0$)
to the conduction band bottom ($E_{\mathrm{F}}=1.17\,$~eV), the
sequence is: $\mathrm{\{Fe_{i}H}\}^{+}+\mathrm{H_{BC}^{+}}$ (net
charge $q=+2$); $\mathrm{\{Fe_{i}H}\}^{0}+\mathrm{H}_{\mathrm{BC}}^{+}$
(net charge $q=+1$); $\mathrm{\{Fe_{i}H}\}^{0}+\mathrm{H}_{\mathrm{T}}^{-}$
(net charge $q=-1$); $\mathrm{\{Fe_{i}H}\}^{-}+\mathrm{\mathrm{H}}_{\mathrm{T}}^{-}$
(net charge $q=-2$). Hence, each kink between adjacent segments corresponds
to a particular transition level identified in the Figure. We note
that transition levels of Fe$_{\mathrm{i}}$ and interstitial H are
measured. The latter is a negative-$U$ defect with donor and acceptor
levels at $E_{\mathrm{c}}-0.18$~eV and $E_{\mathrm{c}}-0.5$~eV,
leading to a $(-/+)$ occupancy level at $E_{\mathrm{c}}-0.34$~eV.\citep{Holm1991,Johnson1995,Bonde99,Bonde2002}
Positive and negatively charged H defects are more stable at the bond-center
($\mathrm{H_{BC}^{+}}$) and tetrahedral interstitial sites ($\mathrm{H_{BC}^{-}}$),
respectively. The neutral charge state (used to calculate the formation
energy in Eq.~\ref{eq:Eform}) was found more stable in the bond-center
site (H$_{\mathrm{BC}}^{0}$). Experimental and calculated levels
are highlighted by open and closed circles on both diagrams. Finally,
on the diagram related to Fe$_{\mathrm{s}}$H$_{n}$ defects, we made
use of color to distinguish A and B structures of Fe$_{\mathrm{s}}$H
and Fe$_{\mathrm{s}}$H$_{2}$ complexes.

Looking at the diagram on the left-hand side of Figure~\ref{fig3}
we conclude that the capture of atomic H by Fe$_{\mathrm{i}}$ is
an energetically favorable process regardless of the Fermi level position.
The energy drop of the formation energy as we move from the upper
segments ($\mathrm{Fe_{i}+2H}$) down to the lower segments ($\mathrm{Fe_{i}H_{2}}$),
represents the binding energy of the reaction $\mathrm{H}+\mathrm{Fe}_{\mathrm{i}}\mathrm{H}_{n}\rightarrow\mathrm{Fe}_{\mathrm{i}}\mathrm{H}_{n+1}$,
with $n=0$ or 1. In intrinsic material (considering the Fermi level
to be approximately at mid-gap), the capture of H by Fe$_{\mathrm{i}}$
corresponds to an energy gain of $\sim0.56\,\mathrm{eV}$. The capture
of a second hydrogen atom corresponds to an energy gain of $\sim0.80\,\mathrm{eV}$
eV, leading to a total binding energy of $\sim1.36\,\mathrm{eV}$
to form a neutral $\mathrm{\mathrm{Fe_{i}H}}_{\mathrm{2}}$ complex.
In \emph{p}-type materials these reactions become less exothermic
and the formation of Fe$_{\mathrm{i}}$H$_{n}$ complexes becomes
less likely. Hydrogenation of Fe$_{\mathrm{i}}$ is further hindered
in $p$-type Si due to the fact that both H and Fe$_{\mathrm{i}}$H$_{n}$
complexes are deep donors, implying a long-range Coulomb repulsion
between reactants. On the other hand, in \emph{n}-type Si there is
no Coulomb barrier for the reaction $\mathrm{Fe_{i}^{0}}+\mathrm{H_{T}^{-}}\rightarrow\mathrm{Fe_{i}H^{-}}$
and the energy drop is $\sim0.6\,\mathrm{eV}$. The reaction with
a second hydrogen atom has a binding energy of 0.75~eV, but it is
likely to be inhibited by repulsion between $\mathrm{H_{T}^{-}}$
and $\{\mathrm{Fe_{s}H}\}^{-}$. In summary, hydrogenation of interstitial
iron in Si leads to Fe$_{\mathrm{i}}$H$_{n}$ complexes whose binding
energies are low, and they are compatible with the the annealing temperature
of 125-175$^{\circ}$C of the FeH-related complex reported in Refs.~\onlinecite{Sadoh1997}
and \onlinecite{Leonard2015}.

The diagram on the right-hand side of Figure~\ref{fig3} immediately
suggests that H binds strongly to Fe$_{\mathrm{s}}$, regardless of
the doping type. In n-type Si the binding energies are $\sim1.4$~eV,
in excellent agreement with the measured binding energy of 1.3~eV
for a FeH-related complex in $n$-type Si, where Fe was suggested
to be at the substitutional site.\citep{Takakashi1999} In the lower
part of the right diagram we also represent the formation energy of
the negatively charged $\{\mathrm{Fe_{s}H_{2}}\}_{\mathrm{B}}^{-}$,
just above the $\mathrm{Fe_{s}H_{2}}(=\!/0)$ negative-$U$ transition
at $E_{\mathrm{c}}-0.26$~eV. This is show as a dashed line to stress
its metastable character.

A more judicious inspection of Figure~\ref{fig3} allows us to conclude
that, while in hydrogen-free material iron impurities have a lower
formation energy at the interstitial site, in H-doped Si, the lower
formation energy complexes are those involving substitutional iron.
This suggests that even in $p$-type Si, high temperature anneals
(maybe with optical excitation in order to avoid Coulomb repulsion
by changing the charge state of either H or Fe impurities), may be
able to convert highly mobile and recombination active Fe$_{\mathrm{i}}$
impurities into stable and low-recombination active Fe$_{\mathrm{s}}$
impurities, where H atoms act as catalysts. Although we are not the
first to realize this possibility,\citep{Szwacki2008} it lacked theoretical
support and it has been overlooked by the solar-Si community.

\subsection{H-assisted diffusivity of iron\label{subsec:diffusion}}

Bearing in mind the observation of the enhancement of Fe-gettering
upon introduction of hydrogen (see for instance Refs.~\onlinecite{Karzel2013,Liu2014,Liu2016}),
and considering the high thermal stability of the species (or phase)
holding the Fe (which survives to temperatures above 500$^{\circ}$C),
we investigated an eventual enhanced migration of $\mathrm{Fe_{i}}$
assisted by H. This could lead to a faster formation of iron precipitates
or out-diffusion from the Si. For the $\mathrm{Fe_{i}}$ we considered
a simple interstitial mechanism through the hexagonal site. In the
case $\mathrm{Fe_{i}H}$, the defect was found to travel as a molecule,
also through neighboring hexagonal sites. As we mentioned before,
a total of 7 NEB images were considered in order to determine the
saddle point along the minimum energy path. The calculated migration
barriers are $0.50\,\mathrm{eV}$, $0.65\,\mathrm{eV}$ and $0.61\,\mathrm{eV}$
for $\mathrm{Fe_{i}^{+}}$, $\mathrm{\{Fe_{i}H\}^{0}}$ and $\mathrm{\{Fe_{i}H\}^{+}}$
respectively. The barrier for migration of $\mathrm{Fe_{i}^{+}}$
is in very good agreement with the measurements, which is about 0.6~eV
(see Ref.~\onlinecite{Istratov1999} and references therein). Both
$\mathrm{\{Fe_{i}H\}^{0}}$ and $\mathrm{\{Fe_{i}H\}^{+}}$ have migration
barriers comparable to that of interstitial iron, and in p-type Si
they are considerably larger than the binding energy of H to Fe$_{\mathrm{i}}$.
These results suggest that hydrogen, if able to attach to Fe$_{\mathrm{i}}$,
does not enhance its diffusivity.

\section{Discussion and conclusions\label{sec:conclusions}}

We calculated the structure, formation energies, binding energies,
and electronic levels of several FeH complexes in Si. $\mathrm{\mathrm{Fe_{i}H}}$
and $\mathrm{\mathrm{Fe_{i}H}}$$_{2}$ defects consist on Fe\nobreakdash-H
and H-Fe-H pseudo-molecules, respectively, with the Fe and H atoms
being located close to hexagonal and tetrahedral interstitial sites
of the lattice. The modest binding energies of the H atoms to Fe$_{\mathrm{i}}$
seem consistent with the annealing temperature in the range of 125-175$^{\circ}$C
reported for a hole trap at $E_{\mathrm{v}}+0.32$~eV and assigned
to an iron-hydrogen complex.\citep{Sadoh1997,Szwacki2008,Leonard2015}
However, an assignment to Fe$_{\mathrm{i}}$H (with a single H atom)
conflicts with its predicted migration barrier, which is close to
that of Fe$_{\mathrm{i}}$. Accordingly, both defects are expected
to anneal out at close temperatures (just above room temperature).
For the same reasons, $\mathrm{\mathrm{Fe_{i}H}}$ complexes are not
able to account for the reduction of $\mathrm{Fe_{i}}$ upon annealing
hydrogenated multicrystalline wafers in the temperature range of 700-900$\lyxmathsym{º}\mathrm{C}$.\citep{Karzel2013,Liu2014,Liu2016}

Fe$\mathrm{_{i}}$H and Fe$\mathrm{_{i}}$H$_{2}$ complexes were
predicted to be simultaneously deep donors and acceptors, and therefore
are not expected to substantially decrease the recombination activity
of Fe in Si. The calculated levels and binding energies suggest that
the donor level measured at $E_{\mathrm{v}}+0.32$~eV from Refs.~\onlinecite{Sadoh1997}
and \onlinecite{Leonard2015} is likely to arise from a Fe$_{\mathrm{i}}$H$_{n}$
complex involving 2 or more H atoms. The Fe$_{\mathrm{i}}$H$_{2}(0/+)$
transition is predicted at $E_{\mathrm{v}}+0.33$~eV, while Fe$_{\mathrm{i}}$H$(0/+)$
is anticipated to occur close to mid-gap.

Substitutional iron and Fe$_{\mathrm{s}}$H$_{n}$ complexes are acceptors.
No donor levels were found for these defects. For the $\mathrm{\mathrm{Fe_{s}H}}$
pair we obtain a single acceptor level close to mid-gap. While this
result is in line with Ref.~\onlinecite{Szwacki2008}, it is not
regarding the calculated barrier for H motion around the Fe$_{\mathrm{s}}$
impurity. In that work, the barrier was estimated to be as low as
0.08~eV, allowing the assignment of neutral $\mathrm{\mathrm{Fe_{s}H}}$
to an isotropic EPR center observed in $n$-type Si.\citep{Takakashi1999}
Our results do not corroborate this view. The calculated barrier for
Fe-H bond reorientation is anticipated to be as high as 0.26~eV,
which is not compatible with a \emph{fast-orbiting} H atom and a motional-averaged
tetrahedral symmetry at $T=10$~K. Further, the near mid-gap location
of the calculated $\mathrm{\mathrm{Fe_{s}H}}(-/0)$ deep acceptor
means that in $n$-type material the stable state is diamagnetic Fe$_{\mathrm{s}}$H$^{-}$
(undetectable by EPR).

Regarding Fe$_{\mathrm{s}}$H$_{2}$, we found a $\langle100\rangle$-aligned
H-Fe-H linear structure in the neutral charge state. The point symmetry
of the defect is $D_{2d}$ and it has an empty doublet in the gap.
Negative and double negative charge states are sensitive to Jahn-Teller
distortions. The negatively charged defect is particularly interesting
as it shows several possible low energy configurations with different
angles between Fe-H bonds, differing by at most 30~meV in their relative
energy and separated by reorientation barriers as shallow as 50~meV.
Based on these findings, we suggest that the FeH-related EPR signal
from Ref.~\onlinecite{Takakashi1999} arises from Fe$_{\mathrm{s}}$H$_{2}^{-}$
or other Fe$_{\mathrm{s}}$H$_{n}$ complex with $n>2$. The later
option is perhaps the most probable as Fe$_{\mathrm{s}}$H$_{2}$
is a negative-$U$ complex with a metastable negative state (see dashed
line segment on the right diagram of Figure~\ref{fig3}).

Figure~\ref{fig3} shows that the formation energy of Fe$_{\mathrm{i}}$
in non-hydrogenated Si is lower than that of Fe$_{\mathrm{s}}$ by
about 0.5~eV. This explains the preference of iron to occupy interstitial
sites. However, in the presence of hydrogen the formation of Fe$_{\mathrm{s}}$-related
complexes becomes favorable. This could explain the formation of large
amounts of Fe$_{\mathrm{s}}$H-related defects in hydrogenated Si,
as detected by EPR after quenching the samples from 950-1250$^{\circ}$C
to 0$^{\circ}$C.\citep{Takakashi1999}

\section*{Acknowledgements}

This work was funded by the Fundação para a Ciência e a Tecnologia
(FCT) under projects PTDC/CTM-ENE/1973/2012 and UID/CTM/50025/2013,
and funded by FEDER funds through the COMPETE 2020 Program. Computer
resources were provided by the Swedish National Infrastructure for
Computing (SNIC) at PDC.

\bibliographystyle{aipnum4-1}

%

\end{document}